# Similarity of symbolic sequences

**B. Kozarzewski**

*University of InformationTechnology and Management*
*H. Sucharskiego 2, 35-225 Rzeszów, Poland*
*bkozarzewski@wsiz.rzeszow.pl*

### Abstract

*A new numerical characterization of symbolic sequences is proposed. The partition of sequence based on Ke and Tong algorithm is a starting point. Algorithm decomposes original sequence into set of distinct subsequences - a patterns. The set of subsequences common for two symbolic sequences (their intersection) is proposed as a measure of similarity between them. The new similarity measure works well for short (of tens letters) sequences and the very long (of hundred thousand letters) as well. When applied to nucleotide or protein sequences may help to trace possible evolutionary of species. As an illustration, similarity of several sets of nucleotide and amino acid sequences is examined.*

## 1. Introduction

Parsing string of symbols from a finite alphabet into substrings is strictly related to time series complexity. One of the first quantitative measure complexity of symbolic sequences has been provided by Lempel and Ziv [1]. They presented an algorithm for parsing string of symbols from a finite alphabet into substrings using copy and insert operations as the following steps. D.-G Ke and Q.-Y. Tong [2] proposed some substantial modification of Lempel-Ziv algorithm by adding replication operation. Recently Kása [3] considered similar parsing of the symbolic sequeces into distinct set of substrings called *d*-substrings. Total number of the *d*-substrings is supposed to be a measure of sequence complexity.

In the following, subsequence resulting from parsing procedure will be call pattern. The pattern is a subsequence with the following properties: it includes an iterative sequence as its prefix, it remembers the history of the sequence and can repeat any series of successive operations in the memory, the last symbol of a lattice must be inserted into lattice unless the end of series is reached. In Appendix one can find listing of corresponding 3-step procedure in the Scilab [4] code. After parsing procedure is completed we are in possesion of a set of distinct patterns which correspond to original sequence, the patterns can be consecutively enumerated. The enumerated set of patterns is called the pattern spectrum *ps* of symbolic sequence.

## 2. Pattern spectrum

Pattern spectrum can be used to identify some properties of symbolic sequence like complexity, entropy and similarity. Here I am going to consider its relation to entropy and similarity measure between two sequences.

1. Entropy

Having spectrum of distinct subsequences in hand it is straightforward to calculate the number of subsequences of given length and afterwards the subsequences length distribution. For example typical sequence of nucleotides shows approximately normal (or may be Weibull like) distribution function with long and very thin tail on the long subsequences side. However spectrum of short amino acids subsequences can be quite different. Let $p(l)$ be probability to find pattern of length $l$ in pattern spectrum, then the Shannon definition of entropy can be used

$$H = -\sum_{l=1}^{l_{max}} p(l) \log_2(p(l)) \qquad (1)$$

2. Similarity between two symbolic sequences

Sequence similarity is an important issue in symbolic sequence alignments. The aim of sequence comparison is to discover similarity relationships between various biological sequences. DNA or proteins can be similar with respect to their sequence of nucleic bases or amino acids . It is belived that high sequence similarity usually implies strong functional or structural similarity. A similarity measure provides a quantitative answer to question whether two sequences show a certain degree of similarity. The distance between two sequences is often used as a numerical measure of how dissimilar the sequences are. The similarity measure is dual to the dissimilarity of the sequences. Lower value of distance means greater similarity. It can also be very usufull to know where the two sequences are similar, and where they differ. An alignment showes similar region (subsequences) between two sequences.

There are numerous measures of similarity between symbolic sequences based on different distance measures between sequence of bases in the gene, or between some structural sequence invariants. The first group of algorithms essentially consists of plotting a point in 4 (or 2)-dimensional space corresponding to a base by moving one unit in a direction associated with the base. The cumulative plot of such points produces a graph that corresponds to the sequence of bases in the gene fragment under consideration. The measure of dissimilarity between two sequences is the Euclidean distance between the end points of the graph. This method has some disadvantages, first of all, the sequences may have different length, second, there is generally no fixed correspondence between their character positions and third, the computational complexity. An alternative to code sequence comparison is Randić [5] approach in which sequences are compared based on a set of structural sequence invariants. The idea is to represent a DNA sequence by an n-dimensional vectors or n-tuple of sequence invariants, typically resulting from a graphical representation of the DNA, and use a distance function for vectors to measure the distance between them. For comparison of very long sequences, e.g. whole genomes, an approach based on the frequency of *k*-mers that appear in a set of sequences are used [6]. A *k*-mer is a series of *k* consecutive letters in a sequence. The approach consists of counting occurrences of *k*-mers in a sequence, for *k* typically ranging from 2 to 8, and apply different statistical methods to *k*-mer distribution. However when the primary sequence is condensed into invariants or *k*-mers there is a loss of information on some aspects of the sequence.

In the presen paper new similarity measure is sugested. With the use of patern spectrum, similarity of two sequences $s_1$ and $s_2$ can be defined simply as

$$sim(s_1, s_2) = \frac{length(int(ps_1, ps_2))}{\sqrt{d(ps_1)d(ps_2)}} \qquad (2)$$

There $int(ps_1, ps_2)$ is a set of sequences common for the two sequences (intersection of $ps_1$ and $ps_2$ ), $length(int(ps_1, ps_2))$ means length of the intersection set and $d(ps) = length(ps)$ is length (dimension) of pattern vector *ps*. Pattern spectra $ps_1$ and $ps_2$ can be different dimension, *sim* value is restricted to the range between null and one. Similarity is symmetric function , so for set of sequences similarity matrix is symmetrical $sim(s_j, s_i) =$ $sim(s_i, s_j)$. Similarity calculation using Eq.2 is no time consuming in contrast to sequence parsing, in particular for sequences as long as $10^6$ or more bases.

To illustrate how the measure introduced works, two short non coding sequences downloaded from [7] are considered. The first sequence is 'n21, AB017710, U50HG, Homo sapiens (human)' of $n = 33$ bases long and the second one is 'n64, AB061824, U58a, Homo sapiens (human)' of $n = 64$ bases long. After parsing is done the patterns spectrum of of first sequence $ps_1$ = ('TAAT', 'CAAT', 'GATGA', 'AAC', 'CTATC', 'CCG', 'AAG', 'CTGATA'), it consists of $d(p_1) = 8$ patterns. There are 3 patterns of 3 bases length, 2 of 4 bases length, 2 of 5 bases length and 1 of 6 bases length. Set of probabilities of pattern lengths is respectively (3/8, 4/8, 5/8, 1/8). Therefore entropy of the first sequence that follows from Eq. (1) is $H = -[3\log_2(3/8) + 4\log_2(2/8) + \log_2(1/8)] = 1.906$. For the second sequence corresponding quantities are $ps_2$ = ('CTGCA', 'GTGA', 'TGACTTTC', 'TTG', 'GGA', 'CACCT', 'TTGG', 'ATTTA', 'CCG', 'TGAAAAT', 'TAAT', 'AAAT', 'TCTG', 'AGCAG'), $d(p_1) = 14$ patterns and their length distribution is (3;3), (5;4), (4,5), (1;7), (1;8), so the entropy of the sequence becomes $H = 2.067$. It seems that entropy value about 2.5 is common for noncoding sequence of bases. Intersection (set of common patterns) of two sequences $int(ps_1, ps_2) = ('CCG', 'TAAT')$ i.e. consists of 2 patterns. Therefore similarity between the two sequences is $sim(s_1, s_2) = 2/\sqrt{8*14} = 0.189$. Besides, common patterns are $p_1$ and $p_7$ from sequence $s_1$ and $p_9$ and $p_{11}$ from sequence $s_2$.

### 3. Comparative study of some sets of base sequences

1. Some 1-exon of β –globin gene sequences

At the beginning a set of exon-1 of the β-globin gene for the 10 species is examined. Sequences data are taken from Liu [8] and Basic [9]. Sequences are rather short, about one hundred bases. Table below lists essential data and entropy for sequences.

|    | Species    | nb  | np | H    |
|----|------------|-----|----|------|
| 1  | Human      | 92  | 20 | 2.48 |
| 2  | Opossum    | 92  | 20 | 2.48 |
| 3  | Gallus     | 92  | 20 | 2.77 |
| 4  | Lemur      | 92  | 21 | 2.77 |
| 5  | Mouse      | 92  | 20 | 2.48 |
| 6  | Rabbit     | 90  | 20 | 2.77 |
| 7  | Gorilla    | 93  | 20 | 2.77 |
| 8  | Bovine     | 86  | 21 | 2.73 |
| 9  | Chimpanzee | 105 | 22 | 2.77 |
| 10 | Goat       | 86  | 20 | 2.73 |

Fig 1 Species name and length of gene, their number of patterns and entropy

*nb* means number of bases, *np* is number of patterns and *H* is entropy of sequence. Note rather low entropy of all sequences.

Similarity numbers between pairs of sequences were calculated from Eq. 2 . Row and column names correspond to species number according to Fig 1. It is seen that, as expected, the (human,gorilla) entry is the largest number, being 0.90, while the smallest one is (lemur,opossum). Similarity = 0 in the case of (lemur,opossum) pair means that there is no common pattern between the two sequences. I have compared my results to those obtained with the method based on 2-dimensional graphical representation of nucleotide sequences and calculation of the distance between geometrical structures. Liu [8] used one of them to characterize dissimilarities for the eight exon–1 β–globin genes.

|    | 1    | 2    | 3    | 4    | 5    | 6    | 7    | 8    | 9    | 10   |
|----|------|------|------|------|------|------|------|------|------|------|
| 1  | 1.00 | 0.20 | 0.30 | 0.15 | 0.45 | 0.50 | 0.90 | 0.54 | 0.86 | 0.45 |
| 2  |      | 1.00 | 0.15 | 0.00 | 0.15 | 0.15 | 0.15 | 0.15 | 0.19 | 0.20 |
| 3  |      |      | 1.00 | 0.20 | 0.15 | 0.25 | 0.30 | 0.29 | 0.29 | 0.25 |
| 4  |      |      |      | 1.00 | 0.20 | 0.20 | 0.15 | 0.29 | 0.19 | 0.15 |
| 5  |      |      |      |      | 1.00 | 0.30 | 0.40 | 0.49 | 0.38 | 0.30 |
| 6  |      |      |      |      |      | 1.00 | 0.50 | 0.39 | 0.48 | 0.30 |
| 7  |      |      |      |      |      |      | 1.00 | 0.54 | 0.91 | 0.45 |
| 8  |      |      |      |      |      |      |      | 1.00 | 0.51 | 0.78 |
| 9  |      |      |      |      |      |      |      |      | 1.00 | 0.43 |
| 10 |      |      |      |      |      |      |      |      |      | 1.00 |

Fig 2 Similarity matrix of the sequences listed in Fig 1

I have transformed his results into similarities and divided entries by factor 4.15 to have the same number for similarity of (human,opossum) pair. First of all, Liu results for all 8 species he considered belong to (0.18,0.22) range, his (lemur,opossum) entry is 0.2 while my is 0.0, his (human,goat) entry is 0.20 while my is 0.45. The exon–1 β–globin genes were also analyzed by Basic [9]. After the same procedure as above his (lemur,opossum) entry is 0.198, his (human,goat) entry is 0.21, his (human,gorilla ) entry is 0.205 while my is 0.90. One can conclude that my results are probably more realistic than that of Liu or Basic ones.

2. Coronavirus complete genomes

The sequence data were retrieved from NCBI GenBank [10]. All of them are medium size, approximately 30000 bases long.

|    | Sequence | nb | np | H |
|----|----------|------|------|------|
| 1  | gi\|12175745\|ref\|NC_002645.1\|Human coronavirus 229E | 27317 | 4316 | 3.13 |
| 2  | gi\|19387576\|ref\|NC_003436.1\|Porcine epidemic diarrhea virus | 28033 | 4491 | 3.86 |
| 3  | gi\|17529670\|gb\|AF220295.1\|Bovine coronavirus strain | 31100 | 4857 | 2.78 |
| 4  | gi\|6625759\|gb\|AF201929.1\|Murine hepatitis virus strain2 | 31276 | 4979 | 3.48 |
| 5  | gi\|9626535\|ref\|NC_001451.1\|Avian infectious bronchitisvirus | 27608 | 4396 | 3.85 |
| 6  | gi\|30275666\|gb\|AY278488.2\|SARS coronavirus BJ01 | 29725 | 4737 | 3.79 |
| 7  | gi\|38304867\|gb\|AY282752.2\|SARS coronavirus CUHK-Su10 | 29736 | 4748 | 3.44 |
| 8  | gi\|30468042\|gb\|AY283794.1\|SARS coronavirus Sin2500 | 29711 | 4749 | 3.95 |
| 9  | gi\|30468045\|gb\|AY283797.1\|SARS coronavirus Sin2748 | 29706 | 4748 | 3.95 |
| 10 | gi\|37361915\|gb\|AY283798.2\|SARS coronavirus Sin2774 | 29711 | 4749 | 3.95 |

Fig 3 Accesion number, name and length of genomes, their number of patterns and entropy

Entropy numbers of all sequences except #3, are significantly higher than that of β –globin gene sequences, in particular sequences #8 to #10.

Similarity data can be compared to those obtained by Wen and Li [11] with the use of alignment-free method. Wen and Li by transforming base sequences into binary sequences found distance matrices of 24 complete coronavirus genomes. The distance matrices were used to characterize dissimilarities coronavirus genes. Like before, I have transformed his results into similarities and divided entries by factor 1.003 to have the same number for similarity of pair (#9,#10) sequences. First of all, Wen and Li lowest results for all 24 ssequences they considered are of order $10^{-2}$ while my are about 0.5. Their (1,2) entry is 0.03 while my is 0.51, their (6,7) entry is 0.99 while my is 0.60, their (8,9) entry and my are the same 0.99. From Fig 4 it follows that all 10 coronavirus sequences can be cassified into four groups: a) sequences #1 to #5, b) sequence #6, c) sequence #7 and d) (probably isoform ) sequences #8 to #10.

|    | 1    | 2    | 3    | 4    | 5    | 6    | 7    | 8    | 9    | 10   |
|----|------|------|------|------|------|------|------|------|------|------|
| 1  | 1.00 | 0.51 | 0.50 | 0.49 | 0.50 | 0.50 | 0.51 | 0.51 | 0.51 | 0.51 |
| 2  |      | 1.00 | 0.50 | 0.50 | 0.51 | 0.50 | 0.51 | 0.50 | 0.50 | 0.51 |
| 3  |      |      | 1.00 | 0.52 | 0.51 | 0.50 | 0.50 | 0.50 | 0.50 | 0.50 |
| 4  |      |      |      | 1.00 | 0.51 | 0.50 | 0.50 | 0.50 | 0.50 | 0.50 |
| 5  |      |      |      |      | 1.00 | 0.50 | 0.50 | 0.50 | 0.50 | 0.50 |
| 6  |      |      |      |      |      | 1.00 | 0.60 | 0.60 | 0.59 | 0.60 |
| 7  |      |      |      |      |      |      | 1.00 | 0.91 | 0.90 | 0.91 |
| 8  |      |      |      |      |      |      |      | 1.00 | 0.99 | 1.00 |
| 9  |      |      |      |      |      |      |      |      | 1.00 | 0.99 |
| 10 |      |      |      |      |      |      |      |      |      | 1.00 |

Fig 4 Similarity matrix of the sequences listed in Fig 3

3. Saccharomyces cerevisiae (yeast) genome

The last set of base sequences analyzed in the present paper are 5 shortest S. cerevisiae chromosomes. The longest one is above $3*10^5$ bases long. Files 2micron.fa, chrI.fa, chrIII.fa, chrVI.fa and chrM.fa were downloaded from [12]

| Chromosom | *nb*   | *np*  | *H*  |
|-----------|--------|-------|------|
| 2micron   | 6318   | 1145  | 1.98 |
| chrI      | 230207 | 30760 | 2.12 |
| chrIII    | 316617 | 41157 | 2.14 |
| chrVI     | 270148 | 35585 | 2.13 |
| chrM      | 85779  | 10312 | 3.18 |

Fig 5 Chromosome name, length, their number of patterns and entropy

Note very low entropy of all but chrM chromosomes, the lowest among all base sequences considered so far in the present paper.

|         | 2micron | chrI | chrIII | chrVI | chrM |
|---------|---------|------|--------|-------|------|
| 2micron | 1.00    | 0.19 | 0.16   | 0.17  | 0.24 |
| chrI    |         | 1.00 | 0.54   | 0.54  | 0.29 |
| chrIII  |         |      | 1.00   | 0.54  | 0.28 |
| chrVI   |         |      |        | 1.00  | 0.29 |
| chrM    |         |      |        |       | 1.00 |

Fig 6 Similarity matrix of the sequences listed in Fig 5

Both, entropy data and similarity data suggest that genomes can be classified into three groups: a) 2micron with *sim* = 0.16 to 0.19 to group b) and *sim* = 0.24 to group c), b) chrI, chrIII and chrVI with *sim* = 0.54 between them and *sim* = 0.28 to 0.29 to group c), and c) chrM.

It can been useful to note that the ratio of patterns sequence $s_1$ shares with sequence $s_2$, it is given by

$$r(s_1) = \frac{length(int(ps_1, ps_2))}{length(ps_1)} = sim(s_1, s_2)\sqrt{\frac{d(ps_2)}{d(ps_1)}} \quad (3)$$

It follows that if for example $s_1$ = 2micro and $s_2$ = chrI then almost 98% of 2micro patterns is present among chrI patterns. It may be also interesting to analyse distribution of pattern lengths. In Fig 7 on the *x*-axis pattern length from 3 to 18 is shown, on the *y*-axis pattern length probability is represented. Typical plot for long nucleotide sequences is that like chrIII.

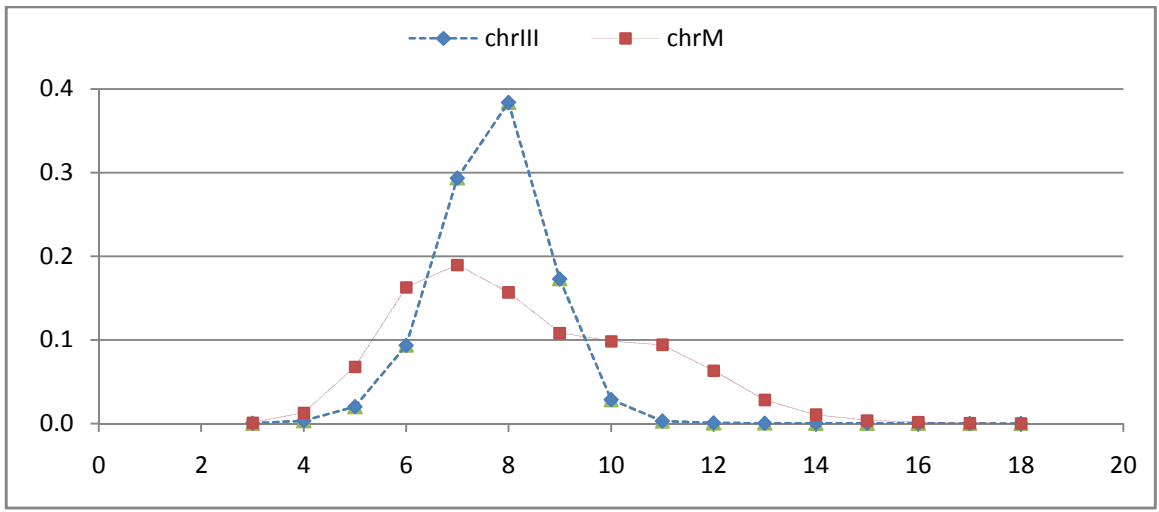

Fig.7 Plot of pattern length probability versus length for chromosomes 2micro and chrIII

The chromosome III results can be approximately reproduced with normal distribution of the mean $\mu$ = 7.7 and standard deviation $\sigma$ = 1.

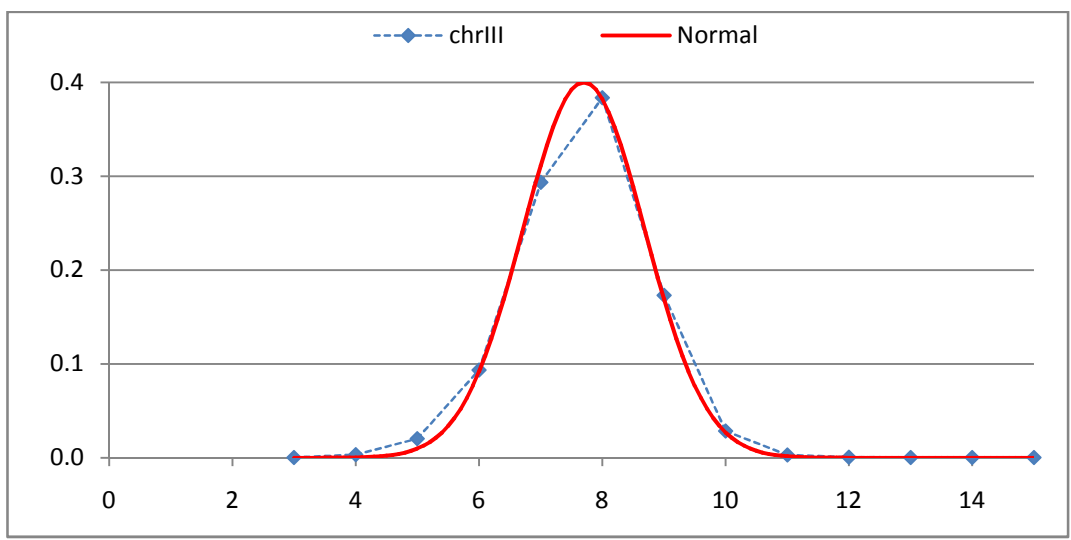

Fig.8 Fit of the chrIII pattern distribution length to the nomal distribution N(7.7,1)

However real sequence distributions exibit long tail. For example, both distributions shown in Fig 7 have patterns as long as 22 bases.

## 4. Some results for amino acid sequences

The rather long sequences of five group of isoform protein data were retrieved from NCBI GenBank file [13].

|    | Protein accesion number and name | naa | np | H |
|----|----------------------------------|-----|-----|------|
| 1  | gi\|126131104\|ref\|NP_733842.2\|fibrocystin isoform 2 [H. sapiens] | 3396 | 493 | 3.23 |
| 2  | gi\|126131102\|ref\|NP_619639.3\|fibrocystin isoform 1 [H. sapiens] | 4073 | 599 | 3.20 |
| 3  | gi\|126131104\|ref\|NP_733842.2\|fibrocystin isoform 2 [H. sapiens] | 3396 | 493 | 3.23 |
| 4  | gi\|113204615\|ref\|NP_000531.2\|ryanodine receptor 1 isoform 1 [H. sapiens] | 5038 | 775 | 3.17 |
| 5  | gi\|113204617\|ref\|NP_001036188.1\|ryanodine receptor 1 isoform 2 [H. sapiens] | 5033 | 774 | 3.17 |
| 6  | gi\|150378539\|ref\|NP_149015.2\|protein piccolo isoform 1 [H. sapiens] | 5142 | 836 | 3.01 |
| 7  | gi\|150170670\|ref\|NP_055325.2\|protein piccolo isoform 2 [H. sapiens] | 4935 | 806 | 3.00 |
| 8  | gi\|325053666\|ref\|NP_001191332.1\|ankyrin-3 isoform 3 [H. sapiens] | 1861 | 277 | 3.16 |
| 9  | gi\|32967601\|ref\|NP_066267.2\|ankyrin-3 isoform 1 [H. sapiens] | 4377 | 660 | 3.12 |
| 10 | gi\|257743025\|ref\|NP_001157980.1\|nebulin isoform 2 [H. sapiens] | 8525 | 1200 | 3.20 |
| 11 | gi\|115527120\|ref\|NP_004534.2\|nebulin isoform 3 [H. sapiens] | 6669 | 958 | 3.18 |
| 12 | gi\|257743023\|ref\|NP_001157979.1\|nebulin isoform 1 [Homosapiens] | 8525 | 1201 | 3.20 |

Fig.9 Protein accesion number and name, amino acid number, number of patterns and entropy

Similarity between different proteins in general is very low. Even a protein isoforms can significantly differ between themselvs like e.g ankyrin-3 isoforms which similarity is as low as 0.58.

|    | 1 | 2 | 3 | 4 | 5 | 6 | 7 | 8 | 9 | 10 | 11 | 12 |
|----|-------|-------|-------|-------|-------|-------|-------|-------|-------|-------|-------|-------|
| 1  | 1.000 | 0.904 | 1.000 | 0.010 | 0.010 | 0.012 | 0.013 | 0.003 | 0.005 | 0.007 | 0.006 | 0.006 |
| 2  |       | 1.000 | 0.904 | 0.013 | 0.013 | 0.013 | 0.013 | 0.002 | 0.006 | 0.007 | 0.007 | 0.007 |
| 3  |       |       | 1.000 | 0.010 | 0.010 | 0.012 | 0.013 | 0.003 | 0.005 | 0.007 | 0.006 | 0.006 |
| 4  |       |       |       | 1.000 | 0.998 | 0.020 | 0.020 | 0.013 | 0.015 | 0.007 | 0.007 | 0.007 |
| 5  |       |       |       |       | 1.000 | 0.020 | 0.020 | 0.013 | 0.015 | 0.007 | 0.007 | 0.007 |
| 6  |       |       |       |       |       | 1.000 | 0.981 | 0.017 | 0.018 | 0.008 | 0.009 | 0.008 |
| 7  |       |       |       |       |       |       | 1.000 | 0.017 | 0.018 | 0.008 | 0.009 | 0.008 |
| 8  |       |       |       |       |       |       |       | 1.000 | 0.582 | 0.007 | 0.006 | 0.007 |
| 9  |       |       |       |       |       |       |       |       | 1.000 | 0.007 | 0.008 | 0.007 |
| 10 |       |       |       |       |       |       |       |       |       | 1.000 | 0.862 | 0.995 |
| 11 |       |       |       |       |       |       |       |       |       |       | 1.000 | 0.867 |
| 12 |       |       |       |       |       |       |       |       |       |       |       | 1.000 |

Fig.10 Similarity matrix of the amino acid sequences listed in Fig 9

Isoform sequences (like that of #1 to #3) show high similarity between themselves but are highly dissimilar to the other proteins of the set analyzed.

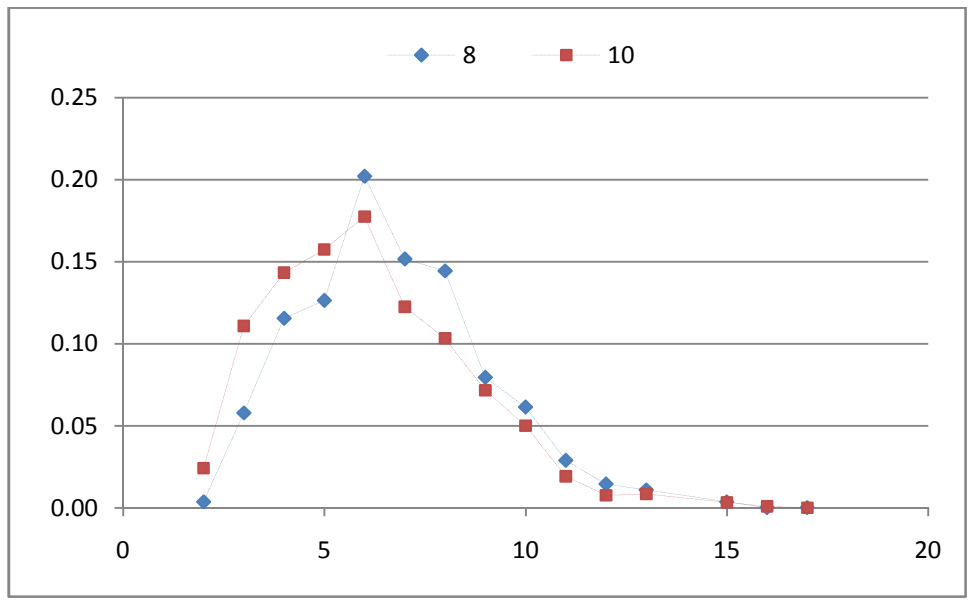

Fig.11 Plot of pattern length probability versus pattern length for protein #6 and #10

## 5. Conclusions

The purpose of the present paper was to examine new measure of entropy of symbolic sequences and similarity between them. Crucial to the approach used is specific sequence parsing algorithm into set of distinct subsquences - patterns. Exploiting length of the set of patterns allows to define complexity, exploiting pattern lengths and its distribution allows to define Shanon entropy of symbolic sequence. Set of common patterns of two symbolic sequences has ben used to define new measure of similarity between them. A possible extension to this work is to investigate complexity of symbolic biological sequences.

**Appendix**

```
// y part of sequence already parsed into patterns, x the rest
//of the sequence
Q = function fL(y,x)
n = size(x,2);
// 1 ------------
Q=x(1);   i=2;
k=strindex(Q,x(i));
while(isempty(k)&(i<n))
Q=Q+x(i); i=i+1;
k=strindex(Q,x(i));end;
// 2 --------------
P=x(i); k=strindex(P,x(i));
while(~isempty(k)&(i<n));
i=i+1; k=strindex(P,x(i));
P=P+x(i); end; Q=Q+P;
//3 -------------
P=Q(1:$-1);
i=size(Q,2); L=[y;P];
k=(L(grep(L,Q))==Q);
while(~isempty(k))
if(i==n), break; end;
i=i+1; Q=Q+x(i);
k=(L(grep(L,Q))==Q;
end; endfunction;
```